\documentclass[aps,prl,twocolumn,superscriptaddress]{revtex4}

\usepackage{epsfig}
\usepackage{amsmath}
\usepackage{amssymb}
\usepackage{textcomp}

\begin{document}
\title{Restricted Dislocation Motion in Crystals of Colloidal Dimer Particles}
\author{Sharon J. Gerbode}
\affiliation{Physics, Cornell University, Ithaca NY 14853}
\author{Stephanie H. Lee}
\affiliation{Materials Science and Engineering, Cornell University, Ithaca NY 14853}
\author{Chekesha M. Liddell}
\affiliation{Materials Science and Engineering, Cornell University, Ithaca NY 14853}
\author{Itai Cohen}
\affiliation{Physics, Cornell University, Ithaca NY 14853}

\begin{abstract}
\label{sec:abstract} \noindent Monolayers of hard dimer colloidal particles consisting of two connected spherical lobes form a degenerate crystal (DC) at high area fractions. In this DC phase, the particle lobes occupy triangular lattice sites while the connections between lobe pairs are randomly oriented, uniformly populating the three crystalline directions of the underlying lattice. In this letter we report on dislocation nucleation and propagation mechanisms observed in DCs and show that certain particle orientations form obstacles blocking dislocation glide. We find that the mean distance between such obstacles is only $\bar{Z}_{exp}=4.6 \pm 0.2$ lattice constants in experimentally observed DC grains. In large simulated DCs with no grain boundaries this average distance is $\bar{Z}_{sim}=6.18 \pm 0.01$ lattice constants, and the probability of finding larger obstacle separations decays exponentially as $\rho(Z)=0.37e^{-Z/4.4}$. Dislocation propagation beyond the obstructions is observed to proceed through dislocation reactions. Assuming that such reactions are the only mechanism used to circumvent these obstacles, we conservatively estimate that the energetic cost of separating a single pair of dislocations in an otherwise defect-free DC grows linearly with the separation. This is in stark contrast to the logarithmically growing separation energy for crystalline monolayers of spheres, hinting that the material properties of DCs may be dramatically different than those observed in crystals of spheres. 
\end{abstract}

\maketitle \setcounter{page}{1} \thispagestyle{empty}

The microscopic motion of dislocations plays a crucial role in melting \cite{Kosterlitz1973,Nelson1979} and governs numerous macroscopic phenomena observed in crystalline materials, including plastic flow, yield, and work hardening \cite{Orowan1934,Polanyi1934,Taylor1934,Cottrell1949}. Studies of dislocations in colloidal crystals enable direct visualization of such processes \cite{Schall2004,Schall2006,Alsayed2005,Zahn1999,Eisenmann2004,Eisenmann2005}, providing an illustrative model for addressing fundamental questions in statistical physics and materials science. Thus far, such studies have focused on crystals of spherical particles, whose defect transport mechanisms are well described by existing models \cite{Orowan1934,Polanyi1934,Taylor1934}. Advances in colloidal particle synthesis techniques have enabled the production of a variety of anisotropic yet monodisperse particles \cite{Manoharan2003,Liddell2003,Johnson2005,Kim2007,Badaire2007,Hernandez2007,Glotzer2007}. Dimer particles are a simple, fundamental extension of spherical particles and can be found in systems ranging from granular piles \cite{Rankenburg2001,Olson2002} to paired adatoms in thin film epitaxy \cite{Qin1997}.  Furthermore, dimers are exceptional since although they are nonspherical, their constituent lobes can nevertheless occupy the lattice sites of crystal structures formed by hard spheres. The study of ordered phases formed by such particles therefore constitutes a natural expansion of the existing body of knowledge on crystals of spheres.

Here we directly examine the mechanisms for dislocation nucleation and propagation in a crystalline phase formed by dense monolayers of colloidal dimer particles.   This crystalline phase, known as a Degenerate Crystal (DC), was first identified in simulations of dimer particles \cite{Woj1991,Woj1992}, and is defined by the following two characteristic features.  First, the individual dimer particle lobes form a triangular lattice; and second, the particle orientations are disordered, uniformly populating the three crystalline directions of the underlying lattice.  We find that dislocation glide in DCs of colloidal dimers is severely limited by geometric constraints formed by certain particle orientations.  This restricted dislocation motion suggests that the material properties of DCs may be dramatically different from those of crystals of spherical particles.

We observe dislocation motion in DCs comprised of hollow, hard dimer particles with spherical lobes of diameter $1.36~\mu$m and lobe separation $1.46~\mu$m. Using sol-gel chemistry, the rhodamine-functionalized silica particles are templated from dimer-shaped hematite cores and are sterically stabilized and suspended in a aqueous solution, yielding nearly hard-core interactions. A detailed description of the particle synthesis is provided in the supplementary materials. The synthesis procedure produces $95\%$ pure dimer particles with particle polydispersity $< 5\%$.  The suspension is pipetted into a sealed wedge-shaped cell, and particle area fraction is controlled by tilting the cell so that particles sediment into the viewing region, which accommodates a monolayer of particles. Before imaging with an inverted microscope, the cell is laid flat, allowing local equilibration until the area fraction is constant over the entire region of interest.  The insertion procedure for filling a wedge cell creates small air bubbles; when these bubbles move near crystalline regions they induce defect formation and transport.

\begin{figure}[ht]
\centering
\includegraphics[height=3in]{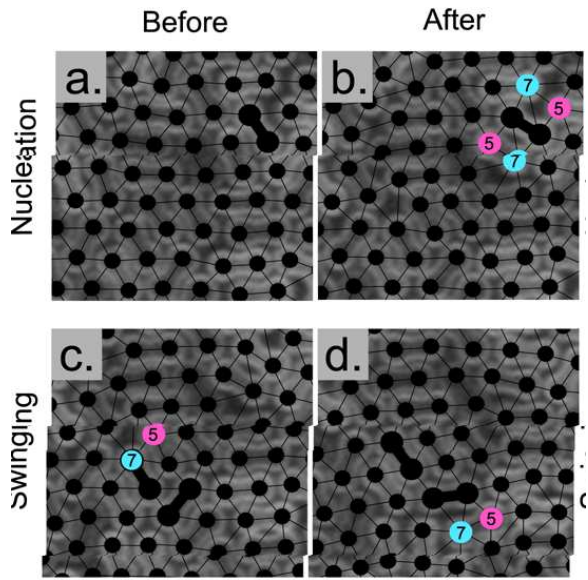}
\caption{(Color online) Before-and-after micrographs illustrating observed dislocation nucleation and glide moves.  Particle lobes have been marked with dots, nearest neighbor bonds are indicated by lines, and dislocations consisting of paired fivefold and sevenfold coordinated lobes have been highlighted.  (a,b) One rotating particle (highlighted with a thick black dumbbell) nucleates a pair of dislocations.  (c,d) A dislocation glides down by three rows through a combination of one sliding move (upper dumbbell) followed by one swinging move (lower dumbbell).}
\label{fig:nucSwingSlide}
\end{figure}

The observed mechanisms for dislocation nucleation and glide are summarized in Fig.~\ref{fig:nucSwingSlide}. Nucleation occurs when a single particle, (highlighted in the image with a thick black dumbbell) rotates, creating a pair of dislocations (Fig.~\ref{fig:nucSwingSlide}a,b). Glide is observed to occur either through a swinging move in which one particle lobe remains fixed while the other swings into a new crystalline position, or via a sliding move in which a particle translates along its axis.  Swinging shifts the dislocation by one crystalline row, while sliding shifts it by two rows.  A sequence of a sliding move followed by a swinging move is shown in Fig.~\ref{fig:nucSwingSlide}c,d, where the sliding and swinging particles have again been highlighted by thick black dumbbells.

\begin{figure}[ht]
\centering
\includegraphics[height=2.15in]{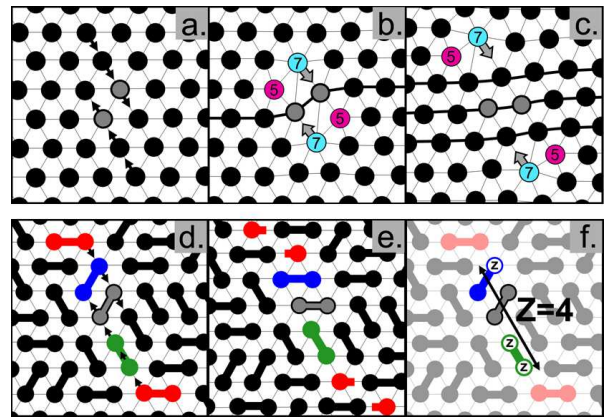}
\caption{Schematic of mechanisms for dislocation nucleation and glide in monolayers of spheres (a - c) and dimer particles (d - f).  (a,b) Displacing two spheres (gray) creates a pair of dislocations, each containing one fivefold and one sevenfold coordinated particle.  (b,c) The pair glides apart when the sevenfold particles shift parallel to their Burgers vectors (outlined arrows) while all other particles retain their crystalline positions. (d) A dislocation pair in a DC is created by rotating one particle (gray) so that its lobes move similarly to the gray spheres in (a). Glide proceeds through the motion of the lobes marked with arrows, either by swinging (blue particle) or sliding (green particle).  (e) Nucleation and glide have the net effect of shifting the left side of the crystal upward and the right side downward. This slip leaves swinging and sliding particles intact, indicating accomodation of the dislocation glide.  The red particles, however, would have to be severed by this deformation.  Since the colloidal particles in our suspensions do not break, the required dislocation motion is blocked by such particle orientations. (f) The sequence of green and blue particles is a zipper of length $Z=4$. This zipper length sets the maximum separation attainable using glide.}
\label{fig:allGlide}
\end{figure}

These observed mechanisms resemble similar mechanisms in crystals of spheres. In such crystals a pair of dislocations is created through the displacement of two adjacent particles (Fig.~\ref{fig:allGlide}a). Each dislocation consists of one fivefold and one sevenfold coordinated particle and is characterized by a Burgers vector. The dislocations glide apart through a succession of moves in which each sevenfold particle shifts its relative lattice position by moving in the direction of the Burgers vector. This process has the net effect of producing slip in the region between the dislocations, shifting the left side of the crystal upward and the right side downward in Fig.~\ref{fig:allGlide}b,c.

Guided by our experimental observations, we note that dislocation motion in DCs is restricted by the constraints of the local particle configuration. A schematic of a DC where dimers are represented by black dumbbells is shown in Fig.~\ref{fig:allGlide}d. Nucleation occurs when a single dimer rotates as shown by the arrows on the gray particle. To allow for glide, the dimer lobes marked with arrows must shift in a manner similar to that shown for spheres. The critical difference is that while spheres are free to move independently, these lobes are constrained to move in collaboration with their partner lobes. The three types of particle orientations relative to the glide direction are illustrated by the green, blue and red particles in Fig.~\ref{fig:allGlide}d. Green particles can shift both lobes in the desired direction by sliding, while blue particles can shift one lobe by swinging. Particles like these therefore permit dislocation glide, in concurrence with the experimental observations of swinging and sliding moves (Fig.~\ref{fig:nucSwingSlide}c,d). The red particles, however, would need to be broken by the relative shifting of the crystal rows during the slip caused by glide (Fig.~\ref{fig:allGlide}d,e). Since the particles in our suspensions do not break, the orientation of such particles blocks the motion of dislocations.  Consequently, only sequences of consecutive swinging and sliding particles allow continuous glide. Since their glide-permitting motion is reminiscent of rearrangements in random square-triangle tilings, we define such sequences as `zippers' \cite{Oxborrow1993}. In Fig.~\ref{fig:allGlide}f, we highlight a single zipper. The zipper lobes, whose motion enables glide, are marked with a `z', and the zipper length, $Z$, is defined as one plus the number of zipper lobes.

\begin{figure}[ht]
\centering
\includegraphics[height=2.55in]{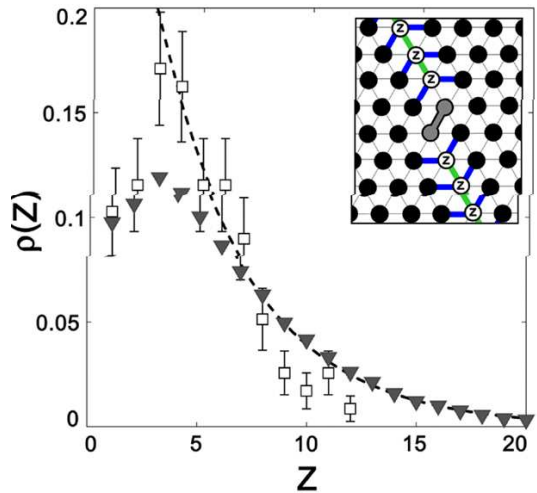}
\caption{(Color online) Probability distribution of zipper lengths in both experimental (empty squares) and simulated (solid triangles) DCs. Counting statistics determine the error bars, which for the simulations are smaller than the symbols. The average zipper length measured from experimental DC grains is $\bar{Z}_{exp}=4.6 \pm 0.2$. Zippers in simulated crystals with $10^4$ lattice sites are only slightly longer: $\bar{Z}_{sim}=6.18 \pm 0.01$. The dotted line is the best fit exponential for $Z>6$, having the form $\rho(Z) = 0.37 e^{-Z/4.4}$. The inset shows the ensemble of glide-permitting particle orientations given nucleation via clockwise rotation of the gray particle.  Particle configurations including a subset of these orientations enable glide via swinging (blue/dark gray) or sliding (green/light gray).}
\label{fig:zipDist}
\end{figure}

In crystals of spheres, dislocations can glide arbitrarily far apart, but in DCs the zipper length defines their maximum glide separation. Consequently, dislocation mobility in DCs is determined by the distribution of zipper lengths. The ensemble of all particle orientations that allow glide for dislocations produced by a clockwise rotation of one particle is shown in the inset of Fig.~\ref{fig:zipDist}. Naively, one might expect to find a zipper of given length with probability $\rho(Z) \propto (2/3)^Z$ since $2/3$ of the particle orientations correspond to swinging or sliding moves. While this crude approximation accurately predicts that long zippers rarely occur, it overlooks important correlations between neighboring particle orientations. 

Precisely accounting for these correlations is theoretically challenging; instead we directly measure $\rho(Z)$ from experimentally observed DCs. Zippers in DC grains are measured by randomly selecting a particle and counting the number of zipper lobes extending from it. We find that, on average, zippers are $\bar{Z}_{exp}=4.6 \pm 0.2$ lattice constants long.~\footnote{For all quantities $\bar{A}$ reported as $\bar{A} \pm \sigma_{\bar{A}}$, the error $\sigma_{\bar{A}}$ represents the standard error of the mean.} The mean diameter of the observed DC grains is $10 \pm 1$ lattice constants. Clearly, $\rho(Z)$ could be affected by this length scale. To determine the zipper length distribution independent of grain size, we prepare ensembles of large DCs using numerical Monte Carlo moves similar to those described in \cite{Woj1992}. The simulations generate crystals with $10^4$ lattice sites, but the mean zipper length is still only $\bar{Z}_{sim}=6.18 \pm 0.01$ (Fig.~\ref{fig:zipDist}). The tail of the simulated distribution is well characterized by the curve $\rho(Z) = 0.37e^{-Z/4.4}$, in agreement with predictions of exponentially decaying orientation correlations for dimers on a triangular lattice \cite{Moessner2001,Krauth2003}.

\begin{figure}[ht]
\centering
\includegraphics[height=1.4in]{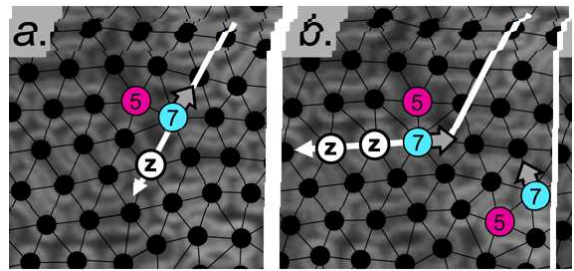}
\caption{(Color online) An observed dislocation reaction allowing a dislocation to hop from one zipper to another.  Only the relevant defects have been highlighted, and their Burgers vectors are indicated by outlined arrows. (a) A dislocation gliding down from the upper right is one lobe from the end of its zipper.  (b) The dislocation has reacted and proceeds by gliding down another zipper extending horizontally to the left. A second dislocation, visible to the lower right of the reaction site, was emitted to conserve total Burgers vector.}
\label{fig:reaction}
\end{figure}

While zippers in DCs are on average only several lattice constants long, shearing or melting processes typically require the transport of dislocations over much larger distances. Our experimental observations reveal a mechanism for surpassing the zipper length limit via dislocation reactions. In such reactions two dislocations may combine or one may split apart so long as the sum of the Burgers vectors is conserved. These reactions allow dislocations to hop onto nearby zippers intersecting the glide path but oriented along a different crystalline axis. An experimentally observed dislocation reaction is illustrated in Fig.~\ref{fig:reaction}. In this sequence, a dislocation gliding down from the upper right approaches the end of its zipper. The defect undergoes a reaction and splits into two new dislocations. One dislocation's Burgers vector is aligned with the horizontal crystalline axis and glides to the left along an available zipper, while the second dislocation has moved to the lower right through a set of moves that are slightly complicated by the presence of a nearby grain boundary (Fig.~\ref{fig:reaction}b).

Such reactions could enable dislocations to separate over arbitrarily large distances along a pathway of intersecting zippers. Nevertheless, the existence of such a pathway does not guarantee that dislocations in DCs are as mobile as those in crystals of spheres. To elucidate the difference between the dislocation transport energetics in the two systems, we compare the cost of separating a single pair of dislocations over $N$ lattice constants along the direction parallel to their Burgers vectors in an otherwise defect-free crystal. In crystals of spherical particles this energy increases as $E_{s}=\frac{\mu a^2}{2\pi (1-\nu)}\text{ln}(N)$, where $a$ is the lattice constant, $\mu$ is the 2-D shear modulus and $\nu$ is the poisson ratio \cite{Weertman1992,Eisenmann2005}. For dislocations separating via intersecting zippers in DCs, each dislocation reaction between zipper segments requires both a core energy to create new defects and a separation energy as one defect glides along the zipper \cite{Weertman1992}.  The energetic cost of separating two dislocations by $Na$ along their Burgers direction using the shortest pathway of connected zipper segments with length $Z_0a$ increases linearly with $N$: $E_{DC}=\frac{\mu a^2}{2\pi (1-\nu)}\text{ln}(Z_0)\left(\frac{4N+Z_0}{5Z_0}\right)$. A detailed calculation of this separation energy is provided in the supplementary matierials. For crooked or fractal-like pathways this energy may increase as an even higher power of $N$.   While dislocation reactions in which two defects merge can release energy, the presence of additional dislocations in the crystal does not guarantee that these would combine with defects produced at zipper junctions, as would be required to lower the energetic cost of separation. Furthermore, even though vacancy-mediated climb could be used to bypass certain obstacles, vacancy transport in DCs can only occur via sliding or swinging particle moves, and consequently dislocation climb is also restricted in DCs.

We claim that the material properties of DCs will be strikingly different from those of crystalline spheres. DCs will be more resistant to plastic flow since dislocation glide cannot proceed along a straight line, as is required for slip. Furthermore, if the separation energy does grow linearly with $N$, we speculate that this will have important implications for melting. In the KTHNY theory of 2-D melting, the crystal to hexatic phase transition requires dislocation pair unbinding \cite{Kosterlitz1973,Nelson1979}. The competition between the energetic cost of dislocation separation and the entropic contribution to the free energy, both of which increase as $\ln(N)$ for crystals of spheres, determines a unique melting temperature. If in DCs the separation energy increases as $N$, dislocation unbinding may no longer be feasible at any finite temperature. This suggests that melting in DCs may occur via additional mechanisms. Furthermore, the observed geometric restrictions in DCs may also apply to other dimer systems, such as lipids with dimer-like head groups \cite{Holz1985} and granular dimers \cite{Rankenburg2001,Olson2002}. For example, these restrictions help explain why avalanches in 2-D piles of dimer beads occur at relatively high critical angles and require tumbling rather than collective slip \cite{Olson2002}. Additional comparative studies between crystals of spheres and DCs should further elucidate how the seemingly benign act of pairing particles into dimers introduces constraints that dramatically alter the material properties of the crystal.

We gratefully thank F. Escobedo, C. Henley, and J. Sethna for reading our manuscript; A. Wolfgang and A. Potter for apparatus development; and P. Chaikin, J. Gregoire, and the I. Cohen research group for stimulating discussion.

\section{Supplementary Materials}

\subsection{Particle synthesis}
The colloidal dimer particles are templated from sacrificial $\alpha$-Fe$_\text{2}$O$_\text{3}$ (hematite) core particles prepared via aging of a condensed Fe(OH)$_\text{3}$ gel as described in \cite{Sugimoto1993}. In a typical synthesis, 50 mL of FeCl$_\text{3}$ (2.0 M) is mixed with 45 mL of NaOH (6.0 M) and 5 mL of the shape-modifier, Na$_\text{2}$SO$_\text{4}$ (0.6 M), before being placed in an oven set at 100\textdegree C for 8 days. A layer of rhodamine isothiocyanate-modified silica is then coated on the $\alpha$-Fe$_\text{2}$O$_\text{3}$ cores using a base catalyzed sol-gel reaction under sonication at 30\textdegree C for 4 hours.  A reaction mixture comprised of 0.4\% hematite powder, 0.25\% rhodamine dye solution, 7.1 M deionized water, 0.92 M ammonia, and 17.4 mM tetraorthoethylsilicate added batch-wise to isopropyl alcohol medium produces fluorescent silica shells approximately 65 nm thick. The shells are stabilized with a non-ionic surfactant, polyvinylpyrrolidone (30000 molecular weight), and finally the hematite cores are selectively etched by dissolution in 18\% hydrochloric acid solution at room temperature. Before use, the colloidal suspension is titrated to pH 7, washed via repeated centrifugation and decanting, and redispersed in deionized water. This synthesis protocal routinely produces $95\%$ pure dimer particles. 

\subsection{Dislocation separation energy calculation}
\begin{figure}[ht]
\centering
\includegraphics[height=3in]{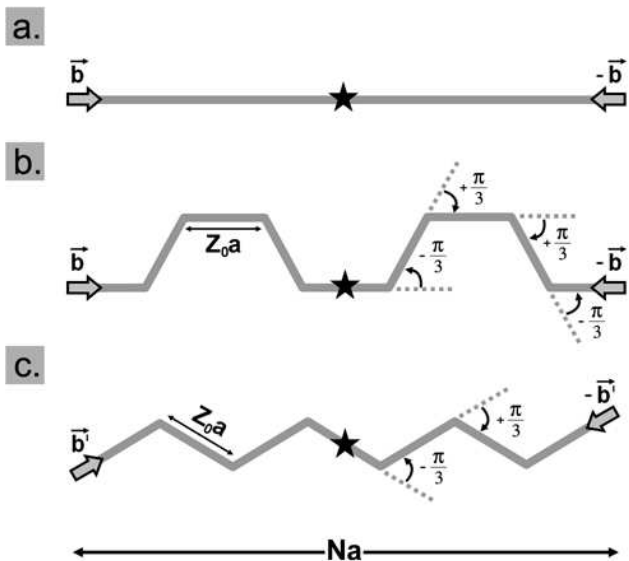}
\caption{Pathways for dislocation pair separation in crystals of spheres (a) and DCs (b,c). Pair nucleation occurs at the location marked with a star in each pathway.  (a) Dislocation pair separation is achieved through nucleation and glide along a straight line parallel to the direction of the defects' burgers vectors $\pm \vec{b}$.  This process results in a pair separation of $Na$. (b) In an otherwise defect-free DC, an identical dislocation pair $\pm \vec{b}$ separates by $Z_0a$, the average zipper length, before undergoing dislocation reactions and traveling along a sequence of tilted path segments of length $Z_0a$. This pathway also results in a separation of $Na$ along the burgers vector direction. (c) The shortest pathway with segment length $Z_0a$ yields a separation of $Na$ along a direction rotated by $\pi /6$ from the axis of the burgers vectors $\pm \vec{b'}$.}
\label{fig:supp}
\end{figure}

\indent In crystals of spheres, nucleated pairs of dislocations with equal and opposite Burgers vectors $\pm \vec{b}$ separate by gliding apart along a straight line parallel to their Burgers vectors (Fig. S\ref{fig:supp}a). This separation allows for relaxation of externally applied shear stresses, and also enables the unbinding of dislocation pairs, which is a crucial component of the KTHNY model of 2-D melting.  Using continuum models of dislocation interactions, the energetic cost of such a separation over $N$ lattice constants is $E_{s}=\frac{\mu a^2}{2\pi (1-\nu)}ln(N)$, where $a$ is the lattice constant, $\mu$ is the 2-D shear modulus and $\nu$ is the poisson ratio \cite{Weertman1992}.  In addition to this continuum interaction, there is also a core energy required to create the dislocations.

Similarly, the cost for separating a pair of dislocations with Burgers vectors $\pm \vec{b}$ in an otherwise defect-free DC can be estimated for any particular intersecting zipper pathway. For example, consider the pathway schematically depicted in Fig. S\ref{fig:supp}b.  In this scenario, two dislocations nucleate and initially glide apart along a straight line parallel to their Burgers vector, mimicking the situation observed in crystals of spheres.  However, after separating by a characteristic distance $Z_0a$, equal to the average zipper length, the defects each undergo dislocation reactions, turning to travel along a new zipper segment of length $Z_0a$ extending along a different crystalline direction. At each reaction site, an extra dislocation is produced, which is assumed to remain stationary.   

The energetic cost of separating the dislocations along the initial straight segment in Fig. S\ref{fig:supp}b is $\frac{\mu a^2}{2\pi (1-\nu)}\text{ln}(Z_0)$. Again, there is also an additional core energy required to create a new dislocation at the junction.  Each additional segment requires more energy, and includes contributions from interactions between all the dislocations along the pathway.  The magnitude of the force between dislocations decreases as the inverse of their separation.  Consequently, the largest force on a dislocation moving on a given segment will come from the dislocation nearest to it.  The energy required for such a separation is  $E_{reac}=\frac{1}{2}\frac{\mu a^2}{2\pi (1-\nu)}\text{ln}(Z_0)$. Adding up these energy contributions from each reaction, as well as the cost of the first straight segment, the total energetic cost for the pathway depicted in Fig. S\ref{fig:supp}b is then

\begin{eqnarray*}
	E_{DC}(Z_0,N) & = & \frac{\mu a^2}{2\pi (1-\nu)}\text{ln}(Z_0) + \frac{8(N-Z_0)}{5Z_0}E_{reac} \\
	& = & \frac{\mu a^2}{2\pi (1-\nu)}\text{ln}(Z_0)\left(\frac{4N+Z_0}{5Z_0}\right).
\end{eqnarray*}

\noindent This separation energy grows linearly with final separation $N$, in contrast to the logarithmically growing energy for dislocation separation in crystals of spheres.  We also note that this is a conservative estimate that excludes the additional core energy costs at each junction.  In the limit that the average zipper length $Z_0$ approaches the separation distance $N$, the energetic cost obtained for crystals of spheres is recovered.

The above energetic cost estimation is only strictly valid for the particular scenario depicted in Fig. S\ref{fig:supp}b.  However, it represents a conservative energetic cost estimate for any pathway spanning $N$ lattice constants along the initial Burgers vector direction, since any other option would require additional segments and dislocation reactions. If the defects are allowed to separate by $Na$ along \emph{any} direction, then the shortest possible pathway is the $-\pi /3$,$+\pi /3$ sequence shown in Fig. S\ref{fig:supp}c.  The energetic cost for separating along this path is $\frac{\mu a^2}{2\pi (1-\nu)}\text{ln}(Z_0)\left(\frac{2N+\sqrt{3}Z_0}{2\sqrt{3}Z_0}\right)$, which still grows linearly with $N$.  More crooked pathways with many more segments could result in an energetic cost of separation that grows as a higher power of $N$. 

We note that in a real system, it is unlikely to find one isolated dislocation pair separating in an otherwise perfect crystal.  The presence of a thermal bath of dislocations may enable other less expensive mechanisms for pair separation.  However, since all dislocation motion is geometrically constrained, the presence of additional dislocations in the crystal does not guarantee that these would be able to combine in merging dislocation reactions, as would be needed to lower the energy of separation.  Future studies of the mechanisms of dislocation unbinding should help elucidate which additional mechanisms can make such separation energetically feasible.

\bibliography{GerbodeRefs}

\end{document}